\documentstyle[proceedings,numreferences]{maxent95}

\def\Re{{\rm I\!R}}
\newcommand{\I}{{\displaystyle{\bf i}}}

\newcommand{\Hs}{\mbox{${\cal H}\ $} }
\newcommand{\R}{\mbox{${\cal R}_{3}\ $}}
\newcommand{\p}{\partial}

\begin{opening}

\title{Are We Cruising a Hypothesis Space?}
\author{C. C. RODRIGUEZ}
\institute{Department of Mathematics and Statistics\\
University at Albany, SUNY\\
Albany NY 12222, USA\\
{\tt carlos@math.albany.edu}}

\end{opening}

\runningtitle{Information Geometry}

\begin{document}
\begin{abstract}

This paper is about Information Geometry, a relatively new subject
within mathematical statistics that attempts to study the problem of
inference by using tools from modern differential geometry. This paper
provides an overview of some of the achievements and possible future
applications of this subject to physics.

\end{abstract}

\section{Introduction}

It is not surprising that geometry should play a fundamental role in
the theory of inference.  The very idea of what constitutes a good
model cannot be stated clearly without reference to geometric concepts
such as size and form of the model as well as distance between
probability distributions. Recall that a statistical model (hypothesis
space) is a collection of probability distributions for the data.
Therefore, a good model should be {\it big} enough to include a {\it
close} approximation to the true distribution of the data, but small
enough to facilitate the task of identifying this approximation.  As
Willy said: {\it as simple as possible but not too simple}.

But there is more. Regular statistical models have a natural
Riemannian structure.  Parameterizations correspond to choices of
coordinate systems and the Fisher information matrix in a given
parameterization provides the metric in the corresponding coordinate
system \cite{kn:mr85,kn:rdrgz89}. By thinking of statistical
models as manifolds, hypothesis spaces become {\it places} and it
doesn't take much to imagine some of these {\it places} as models for
the only physical place there is out there, namely: {\bf
spacetime}. In section~\ref{sec:radially} of the paper we apply the
techniques of information geometry to show that the space of radially
symmetric distributions admits a foliation into pseudo-spheres of
increasing radius. If we think of a radially symmetric distribution as
describing an uncertain physical position we discover a hypothesis
space that, in many ways resembles an expanding spacetime. An
isotropic, homogeneous space with pseudo-spherical symmetries and with
{\it time} increasing with decreasing curvature radius. This
admittedly simple toy model of reality already suggests a number of
truly remarkable consequences for the nature of spacetime.  Here are
three of them:
\begin{enumerate}
\item 
The appearance of time is a consequence of uncertainty.  
\item
Space is infinite dimensional and only {\it on the average} appears as
four dimensional. 
\item
Spin is a property of space and not of a particle so that all truly
fundamental particles must have spin.
\end{enumerate}

I must emphasize that, at the time of writing, there is no direct
experimental evidence in favor of any of the above statements.
Nevertheless there is indirect evidence that they should not be too
quickly dismissed as nonsense.  

With respect to the first statement.  Recall that the appearance of
the axis of time, in standard general relativity, is a consequence of
specifying an initial and a final 3-geometry on two spacelike
hypersurfaces plus evolution according to the field
equation~\cite{baierlein62}.  Time is therefore a consequence of
3-space geometry and the field equation of general relativity, which
in turn seems to be of an statistical nature (see \cite{jacobson95}
and section~\ref{sec:curvature} below).  These, I believe, are facts
that support, at least in spirit, the first claim above.

There is absolutely no evidence that space has infinitely many
dimensions, but had this be true, it would explain why we observe only
four of them.  It also seems a priori desirable to have a model that
produces {\it observed space} as a macroscopic object not unlike
pressure or temperature.

With respect to the third statement.  Hestenes\cite{hestenes66} shows
that many of the rules for manipulating spin have nothing to do with
quantum mechanics but are just general expressions for the geometry of
space. It is also worth noticing that the standard model allows the
existence of elementary particles without spin but these have not yet
been observed.

But there is still more. Think of the different roles that the
concepts of, entropy, curvature, and, local hyperbolicity, play in
statistics and in physics and you will realize that the link is a
useful bridge for transporting ideas from physics to statistics and
vice-versa. The following sections (\ref{sec:entropy},
\ref{sec:curvature}, \ref{sec:hyperbolicity}) of this paper do exactly
that.  That is, they examine the meaning of each of these concepts
(entropy, curvature and local hyperbolicity) keeping the proposed link
between inference and physics in mind.

The link between information geometry and general relativity promises
big rewards for both, statistical inference and physics. For example,
statisticians may look at the field equations of general relativity as
a procedure for generating statistical models from prior information
encoded in the distribution of energy and matter fields. On the other
hand, physicists may see information geometry as a possible language
for the elusive theory of quantum gravity since it is a language
already made out of the right ingredients: {\it uncertainty} and {\it
differential geometry}.

\section{The Hypothesis Space of Radially Symmetric Distributions} 
\label{sec:radially}

Let \R be the collection of all radially symmetric distributions of
three dimensional euclidean space. The probability of an infinitesimal
region around the point $x\in \Re^{3}$, of volume $d^{3}x$, that is
assigned by a general element of \R is given by,

\begin{equation}
P( d^{3}x |\psi,\theta,\sigma ) =  \frac{1}{\sigma^{3}}
\left| \psi\left( \left|\frac{x-\theta}{\sigma}\right|^{2} 
	\right)\right|^{2} d^{3}x	\label{eq:d3x}
\end{equation}
where, $\theta \in \Re^{3}$ is a location parameter, $\sigma > 0$ is a
scale parameter and $\psi$ is an arbitrary differentiable function of
$r^{2}>0$ satisfying the normalization condition:

\begin{equation}
\int_{0}^{\infty} r^{2}|\psi(r^{2})|^{2} dr = \frac{1}{4\pi}
		\label{eq:4pi}
\end{equation}
Equation (\ref{eq:4pi}) assures that the probability assigned to the
whole space by (\ref{eq:d3x}) is in fact 1. The derivative,
$\psi^{'}$, must also decrease to 0 sufficiently fast so that the
integrals (\ref{eq:gmn}) exist.  Since $\psi$ is an infinite
dimensional parameter, \R is also an infinite dimensional manifold but
the space , $\R(\psi)$, of radially symmetric distributions for a
given function $\psi$ is a four dimensional submanifold of $\R$
parameterized by $(\theta_{0},\theta_{1},\theta_{2},\theta_{3}) =
(\sigma,\theta)$.  The metric in $\R(\psi)$ is given by the $4 \times
4$ Fisher information matrix (see \cite{kn:brgmv.81} p. 63) with entries:

\begin{equation}
g_{\mu\nu} = 4 \int (\p_{\mu}f)(\p_{\nu}f) d^{3}x
		\label{eq:gmn}
\end{equation}
where $\mu,\nu = 0,\ldots,3$, the function $f$ is the square root of
the density given in (\ref{eq:d3x}) i.e.,

\begin{equation}
f(x|\theta,\sigma) = \sigma^{-3/2} 
\psi\left( \left|\frac{x-\theta}{\sigma}\right|^{2} \right)
		\label{eq:fx}
\end{equation}
and $\p_{\mu}$ denotes the derivative with respect to $\theta_{\mu}$.
Let us separate the computation of the metric tensor terms into three
parts. The entries $g_{ij}$, the entries $g_{0i}$ for $i,j=1,2,3$ and
the element $g_{00}$. Replacing (\ref{eq:fx}) into (\ref{eq:gmn}),
doing the change of variables $x=\theta+\sigma y$ and using the fact
that 
\begin{equation}
\p_{i}\psi(y^2) = -2 y_{i}\psi^{'}(y^2)/\sigma \label{eq:pi}
\end{equation}
we get,
\begin{equation}
g_{ij} = \frac{16}{\sigma^{2}}
 \int y_{i} y_{j}\left(\psi^{'}(y^{2})\right)^{2} d^{3}y \label{eq:gij}
\end{equation}
where $y^{2}=|y|^{2}$ is the Clifford product of the vector $y$ by
itself. Carrying out the integration in spherical coordinates we obtain,
\begin{equation}
g_{ij} = 0 \mbox{\ \ for $i \ne j$} \label{eq:neij}
\end{equation}
and,
\begin{equation}
g_{ii} = \frac{64\pi}{3\sigma^{2}}\int r^{4}|\psi^{'}(r^{2})|^{2} dr
	\label{eq:gii}
\end{equation}
The derivative with respect to $\sigma$ of the function given
in (\ref{eq:fx}) is,

\begin{equation}
\p_{0}f = -\frac{\sigma^{-5/2}}{2}
	\left[ 3\psi + 4 y^{2}\psi^{'} \right] \label{eq:p0f}
\end{equation}
and therefore from this and (\ref{eq:pi}) we have,
\begin{equation}
g_{0i} = 4 \int (\p_{0}f)(\p_{i}f) d^{3}x
	\propto \int [3\psi + 4y^{2}\psi^{'}] y_{i}\psi^{'} d^{3}y = 0
	\label{eq:g0i}
\end{equation}
where the value of 0 for the last integral follows by performing the
integration in spherical coordinates or simply by symmetry, after
noticing that the integrand is odd. Finally from (\ref{eq:p0f})
we get,

\begin{equation}
g_{00} = \frac{4\pi}{\sigma^{2}} \int [ 3\psi(r^{2}) +  
	4r^{2}\psi^{'}(r^{2}) ]^{2} r^{2} dr	\label{eq:g000}
\end{equation}
Expanding the square and integrating the cross term by parts to
show that,

\begin{equation}
\int r^{4}\psi(r^{2})\psi^{'}(r^{2}) dr = -\frac{3}{4}(\frac{1}{4\pi})
	\label{eq:r4}
\end{equation}
where we took $u=\psi r^{3}/2$ and $v^{'} = 2r\psi^{'}$ for the 
integration by parts and we have used (\ref{eq:4pi}). We obtain,

\begin{equation}
g_{00} = \frac{4\pi}{\sigma^{2}}\left[ \frac{-9}{4\pi} + 
	16\int r^{6} |\psi^{'}(r^{2})|^{2} dr \right]
	\label{eq:g00}
\end{equation}
The full matrix tensor looks like this,

\begin{equation}
(g) = \frac{1}{\sigma^{2}}
\left [\begin {array}{cccc} {\it J(\psi)}&0&0&0\\\noalign{\medskip}0&{\it 
K(\psi)}&0&0\\\noalign{\medskip}0&0&{\it K(\psi)}&0\\\noalign{\medskip}0&0&0&{
\it K(\psi)}\end {array}\right ]
	\label{eq:(g)}
\end{equation}
where $J(\psi)$ and $K(\psi)$ are just short hand notations for the
factors of $\frac{1}{\sigma^{2}}$ in (\ref{eq:g00}) and
(\ref{eq:gii}).  These functions are always positive and they depend
only on $\psi$. Straight forward calculations, best done with a
symbolic manipulator like MAPLE, show that a space with this metric
has constant negative scalar curvature given by $-1/J(\psi)$. It 
follows that for a fix value of the function $\psi$ the hypothesis
space of radially symmetric distributions $\R(\psi)$ is the 
pseudo-sphere of radius $J^{1/2}(\psi)$. We have therefore
shown that the space of radially symmetric distributions has a
foliation (i.e. a partition of submanifolds) of pseudo-spheres of
increasing radius.

This is a mathematical theorem. There can be nothing controversial about it.
What it may be disputed, however, is my belief that the hypothesis
space of radially symmetric distributions may be telling us something new
about the nature of real physical spacetime. What I find interesting is
the fact that if we think of position subject to radially symmetric
uncertainty then the mathematical object describing the positions (i.e.
the space of its distributions) has all the symmetries of space plus time.
It seems that time, or something like time, pops out automatically when
we have uncertain positions. I like to state this hypothesis with
the phrase: 
{\center {\it there is no time, only uncertainty}\\ }

\subsection{Uncertain Spinning Space?}
The hypothesis space of radially symmetric distributions is the
space of distributions for a random vector $y\in \Re^{3}$ of 
the form,
\begin{equation}
y = x +  \epsilon	\label{eq:y=x+}
\end{equation}
where $x\in\Re^{3}$ is a non random location vector, and
$\epsilon\in\Re^{3}$ is a random vector with a distribution radially
symmetric about the origin and with standard deviation $\sigma > 0$ in
all directions.  It turns out that exactly the same hypothesis space
is obtained if instead of (\ref{eq:y=x+}) we use,
\begin{equation}
y = x + \I\ \epsilon	\label{eq:y=x}
\end{equation}
where $\I$ is the constant unit pseudo scalar of the Clifford algebra
of $\Re^{3}$. The pseudo scalar $\I$ has unit magnitude, commutes with
all the elements of the algebra, squares to $-1$ and it represents the
oriented unit volume of $\Re^{3}$ \cite{hestenes84}. By taking
expectations with the probability measure indexed by $(x,\sigma,\psi)$
we obtain that,
\begin{equation}
E(y|x,\sigma,\psi) = x	\label{eq:Ey}
\end{equation}
and,
\begin{equation}
E(y^{2}|x,\sigma,\psi) = x^{2} - \sigma^{2}	\label{eq:Ey2}
\end{equation}
Equation (\ref{eq:Ey2}) shows that, even though the space of radially
symmetric distributions is infinite dimensional, on the average
the intervals look like the usual spacetime intervals.

We may think of $y$ in (\ref{eq:y=x}) as encoding a position in 3-space
together with an uncertain degree of orientation given by the bivector
part of $y$, i.e. $\I\epsilon$. In other words we assign to the point
$x$ and intrinsic orientation of direction $\hat{\epsilon}$ and 
magnitude $|\epsilon|$. In this model the uncertainty is not directly
about the location $x$ (as in (\ref{eq:y=x+})) but about its postulated
degree of orientation (or spinning).

\section{Entropy and Ignorance} \label{sec:entropy}

The notion of statistical entropy is not only related to the
corresponding notion in physics it is exactly the same thing as
demonstrated long ago by Jaynes \cite{jaynes57}.  Entropy appears
indisputable as the central quantity of information geometry.  In
particular, from the Kullback number (relative entropy) between two
distributions in the model we obtain the metric, the volume element, a
large class of connections \cite{kn:rdrgz89}, and a notion of
ignorance within the model given by the so called entropic priors
\cite{rodriguez91}.  In this section I present a simple argument,
inspired by the work of Zellner on MDIP priors \cite{zellner95},
showing that entropic priors are the statistical representation of the
vacuum of information in a given hypothesis space.

Let $\Hs = \{ f(x|\theta): \theta \in \Theta \}$ be a general regular
hypothesis space of probability density functions $f(x|\theta)$ for a
vector of observations $x$ conditional on a vector of parameters
$\theta = (\theta^{\mu})$.  Let us denote by $f(x,\theta)$, the joint
density of $x$ and $\theta$ and by $f(x)$ and $\pi(\theta)$ the
marginal density of $x$ and the prior on $\theta$ respectively.  We
have,
\begin{equation}
f(x,\theta) = f(x|\theta) \pi(\theta)	\label{eq:fxt}
\end{equation}
Since \Hs is regular, the Fisher information matrix,
\begin{equation}
g_{\mu\nu}(\theta) = 4 \int (\p_{\mu}f^{1/2})(\p_{\nu}f^{1/2}) dx
		\label{eq:gmnt}
\end{equation}
exists and it is continuous and positive definite (thus non singular)
at every $\theta$. As in (\ref{eq:gmn}), $\p_{\mu}$ denotes the
partial derivative with respect to $\theta^{\mu}$. The space \Hs with
the metric $g = (g_{\mu\nu})$ given in (\ref{eq:gmnt}) forms a
Riemannian manifold. Therefore, the invariant element of volume is
given by,
\begin{equation}
\eta(d\theta) \propto \sqrt{\det\  g(\theta)} d\theta	\label{eq:eta}
\end{equation}
This is in fact a differential form \cite[p. 166]{kn:dbrvn.84} that
provides a notion of surface area for the manifold \Hs and it is
naturally interpreted as the uniform distribution over \Hs.  Formula
(\ref{eq:eta}), known as Jeffeys rule, is often used as a universal
method for building total ignorance priors.  However, (\ref{eq:eta})
does not take into account the fact that a truly ignorant prior for
$\theta$ should contain as little information as possible about the
data $x$.  The entropic prior in \Hs demands that the joint
distribution of $x$ and $\theta$, $f(x,\theta)$, be as difficult as
possible to discriminate from the independent model $h(x) \sqrt{\det\
g(\theta)}$, where $h(x)$ is an initial guess for $f(x)$.  That is, we
are looking for the prior that minimizes the Kullback number between
$f(x,\theta)$ and the independent model, or in other words, the
prior that makes the joint distribution of $x$ and $\theta$ to have
maximum entropy relative to the measure $h(x)\sqrt{\det g(\theta)}dx
d\theta$.  Thus, the entropic prior is the density $\pi(\theta)$ that
solves the variational problem,
\begin{equation}
\min_{\displaystyle\pi}
	\int f(x,\theta) \log 
	\frac{f(x,\theta)}{h(x)\sqrt{\det\ g(\theta)}}\ dx\ d\theta
	\label{eq:min}
\end{equation}
Replacing (\ref{eq:fxt}) into (\ref{eq:min}), simplifying, and using a
lagrange multiplier, $\lambda$, for the normalization constraint, that $\int
\pi(\theta)d\theta = 1$, we find that $\pi$ must minimize,
\begin{equation}
\int \pi(\theta) I(\theta:h)\ d\theta + 
  \int \pi(\theta) \log \frac{\pi(\theta)}{\sqrt{\det g(\theta)}}\ d\theta
    + \lambda \int \pi(\theta)\ d\theta	\label{eq:L}
\end{equation}
where, $I(\theta:h)$ denotes the Kullback number between $f(x|\theta)$ and
$h(x)$, i.e.,
\begin{equation}
I(\theta:h) = \int f(x|\theta) \log \frac{f(x|\theta)}{h(x)}\ d\theta
	\label{eq:Ith}
\end{equation}
The Lagrangian ${\cal L}$ is given by the sum of the integrands in
(\ref{eq:L}) and the Euler-Lagrange equation is then,
\begin{equation}
\frac{\p{\cal L}}{\p\pi} = I(\theta:h) + 
  \log \frac{\pi(\theta)}{\sqrt{\det g(\theta)}} + 1 + \lambda = 0
	\label{eq:ELE}
\end{equation}
from where we obtain that,
\begin{equation}
\pi(\theta) \propto e^{{\displaystyle -I(\theta:h)}} 
	\sqrt{\det g(\theta)}	\label{eq:pitp}
\end{equation}
The numerical values of the probabilities obtained with the formula
(\ref{eq:pitp}) depend on the basis for the logarithm used in
(\ref{eq:min}).  However, the basis for the logarithm that appears in the
definition of the Kullback number is arbitrary (entropy is defined
only up to a proportionality constant).  Thus, (\ref{eq:pitp}) is not
just one density, but a family of densities,
\begin{equation}
\pi(\theta|\alpha,h) \propto e^{{\displaystyle -\alpha I(\theta:h)}} 
	\sqrt{\det g(\theta)}	\label{eq:pita}
\end{equation}
indexed by the parameter $\alpha > 0$ and the function $h$.  Equation
(\ref{eq:pita}) is the family of entropic priors introduced in
\cite{kn:rdrgz89} and studied in more detail in
\cite{kn:rdrgz90},\cite{rodriguez91} and \cite{rodriguez93}.

It was shown in \cite{rodriguez91} that the parameter $\alpha$ should
be interpreted as the number of virtual observations supporting 
$h(x)$ as a guess for the distribution of $x$. Large values
of $\alpha$ should go with reliable guesses for $h(x)$ but, as it 
was shown in \cite{rodriguez93}, the inferences are less robust. This
indicates  that ignorant priors should be entropic priors with the
smallest possible value for $\alpha$, i.e., with,
\begin{equation}
\alpha^{*} = \inf\{\alpha > 0 : \int 
e^{\displaystyle -\alpha I(\theta,h)} \ \eta(d\theta) < \infty\}
	\label{eq:alphastar}
\end{equation}
Here is the canonical example.

\subsection{Example: The Gaussians}

Consider the hypothesis space of one dimensional gaussians parameterized by
the mean $\mu$ and the standard deviation $\sigma$.
When $h$ is an arbitrary gaussian with parameters $\mu_{0}$ and $\sigma_{0}$
straight forward computations show that the entropic prior is given by,
\begin{equation}
\pi(\mu,\sigma|\alpha,\mu_{0},\sigma_{0}) = \frac{1}{Z} {\sigma}^{\alpha-2}
\exp\left[{{\frac {-\alpha\,\left (\left (\mu-\mu_{0}\right )^{2}
+{\sigma}^{2}\right )}{2\sigma_{0}^{2}}}}\right]
	\label{eq:pims}
\end{equation}
where the normalization constant $Z$ is defined for $\alpha > 1$
and is given by,
\begin{equation}
Z = \frac{2}{\sqrt{\pi}}\left(\frac{\alpha}{2}\right)^{\alpha/2}
\Gamma (\frac{\alpha - 1}{2})^{-1} {\sigma_{0}}^{-\alpha}  \label{eq:C}
\end{equation}
Thus, in this case $\alpha^{*} = 1$ and the most ignorant prior is
obtained by taking the limit $\alpha \rightarrow 1$ and $\sigma_{0}
\rightarrow \infty$ in (\ref{eq:pims}) obtaining, in the limit, an
improper density proportional to $1/\sigma$, which makes every body
happy, frequentists and bayesians alike.

\section{Curvature and Information} \label{sec:curvature}

Curvature seems to be well understood only in physics, specially from
the point of view of gauge theories where the curvature form
associated to a connection has been shown to encode field strengths
for all the four fundamental forces of nature \cite{bleecker81}.  In
statistics, on the other hand, the only thing we know (so far) about
the role of curvature is that the higher the scalar curvature is at a
given point of the model, the more difficult it is to do estimation at
that point.  This already agrees nicely with the idea of black holes ,
for if in a given model there is a curvature $R_{0}$ beyond which
estimation is essentially impossible then the space is partitioned
into three regions with curvatures, $R < R_{0}$, $R = R_{0}$ and $R >
R_{0}$ that correspond to regular points, horizon points and points
inside black holes.  No body has found an example of a hypothesis
space with this kind of inferential black holes yet, but no body has
tried to look for one either.  Before rushing into a hunt it seems
necessary to clarify what exactly it is meant by the words: {\it
estimation is essentially impossible at a point}.

I believe that one of the most promising areas for research in the
field of information geometry is the clarification of the role of
curvature in statistical inference. If indeed physical spacetime can
be best modeled as a hypothesis space then, what is to be learned from
the research on statistical curvature will have direct implications
for the nature of physical space. On the other hand, it also seems
promising to re-evaluate what is already known in physics about
curvature under the light of the proposed link with inference.  Even a
naive first look will show indications of what to expect for the role
of curvature in inference. Here is an attempt at that first look.
\begin{description}

\item[From the classic statement:] {\em Mass-energy is the source of
gravity and the strength of the gravity field is measured by the
curvature of spacetime }

\item[We guess:] {\em Information is the source of the curvature of
hypothesis spaces. That is, prior information is the source of the form of
the model}

\item[From:] {\em The dynamics of how mass-energy curves spacetime are
controlled by the field equation:}
\begin{equation}
{\bf G} = \kappa {\bf T}
\end{equation}
{\em where {\bf G} is the Einstein tensor, {\bf T} is the
stress-energy tensor and $\kappa$ is a proportionality factor}

\item[guess:] {\em The field equation controls the dynamics of how
prior information produces models}

\item[From:] {\em The field equation for empty space is the
Euler-Lagrange equation that characterizes the extremum of the Hilbert
action, with respect to the choice of geometry. That is it extremizes}
\begin{equation}
S_{g} = \int R d\Omega, \ \ \ \ d\Omega = \sqrt{|\det g|} d^{4}x,
\end{equation}
{\em where the integral is taken over the interior of a
four-dimensional region $\Omega$, $R$ is the scalar curvature and $g$
is the metric}

\item[guess1:] {\em The form of hypothesis spaces based on no prior
information must satisfy}
\begin{equation}
R_{ij} - \frac{1}{2}R g_{ij} = 0 \label{eq:Rij}
\end{equation}
{\em where $g_{ij}$ is the Fisher information matrix, $R_{ij}$ is the
Ricci tensor and $R$ is the scalar curvature as above.}

\item[guess2:] {\em Given a hypothesis space with Fisher information
matrix $g(\theta)$, the Einstein tensor, $G$, i.e. the left hand side
of (\ref{eq:Rij}), quantifies the amount of prior information locally
contained in the model at each point $\theta$.}

\end{description}

\section{Hyperbolicity}
\label{sec:hyperbolicity}

What it seems most intriguing with respect to the link between
information geometry and general relativity is the role of
hyperbolicity.  We know from general relativity that physical
spacetimes are Riemannian manifolds which are locally Lorentzian.
That is, at each point, the space looks locally like Minkowski space.
Or, in other words, the symmetries of the tangent space at each point
are those of hyperbolic space.  On the other hand, in information
geometry, hyperbolicity appears at two very basic levels.  First,
hyperbolicity appears connected to the notion of regularity through
the property of {\it local asymptotic normality} (LAN for short see
\cite{kn:brgmv.81}). This is in close agreement with what happens in
physics.  The LAN property says that the manifold of distributions of
$n$ independent and identically regularly distributed observations can
be locally approximated by gaussians for large n, and since the
gaussians are known to form hyperbolic spaces, the correspondence with
physics is perfect.  Second, in statistical inference hyperbolicity
also appears mysteriously connected to entropy and Bayes' theorem!
(see my {\it From Euclid to Entropy}\cite{kn:rdrgz91}) and by
following the link back to general relativity we obtain a completely
new and unexpected result: {\it entropy and Bayes theorem are the
source of the local hyperbolicity of spacetime!}.  That entropy and
thermodynamics are related to general relativity may have seem
outrageous in the past, but not today. It does not seem outrageous at
all when we consider that, Bekenstein found that the entropy of a
black hole is proportional to its surface area \cite{bekenstein73},
when we consider that Hawking discovered that black holes have a
temperature \cite{hawking75} and specially when we consider that
Jacobson showed that the field equation is like an equation of state
in thermodynamics \cite{jacobson95}.

\bibliography{carlos}
\bibliographystyle{maxent95}

\end{document}